\documentclass[aps,prl,showpacs,preprintnumbers,nofootinbib,floatfix,twocolumn]{revtex4}

\usepackage{bm}
\usepackage{latexsym}
\usepackage{dcolumn}
\usepackage{amsfonts,amssymb}
\usepackage{graphicx,epsfig}
\usepackage{psfrag}

\def \lleq {\lower0.9ex\hbox{ $\buildrel < \over \sim$} ~}
\def \ggeq {\lower0.9ex\hbox{ $\buildrel > \over \sim$} ~}

\def \beq  {\begin{equation}}
\def \eeq  {\end{equation}}
\def \ber  {\begin{eqnarray}}
\def \eer  {\end{eqnarray}}

\begin{document}
\newcommand{\newc}{\newcommand}

\newcommand{\ben}{\begin{eqnarray}}
\newcommand{\een}{\end{eqnarray}}
\newc{\be}{\begin{equation}}
\newc{\ee}{\end{equation}}
\newc{\ba}{\begin{eqnarray}}
\newc{\ea}{\end{eqnarray}}
\newc{\bea}{\begin{eqnarray*}}
\newc{\eea}{\end{eqnarray*}}
\newc{\D}{\partial}
\newc{\ie}{{\it i.e.} }
\newc{\eg}{{\it e.g.} }
\newc{\etc}{{\it etc.} }
\newc{\etal}{{\it et al.}}
\newcommand{\nn}{\nonumber}
\newc{\ra}{\rightarrow}
\newc{\lra}{\leftrightarrow}
\newc{\lsim}{\buildrel{<}\over{\sim}}
\newc{\gsim}{\buildrel{>}\over{\sim}}
\title{Vacuum Energy, the Cosmological Constant and Compact Extra Dimensions: \\Constraints from Casimir Effect Experiments}
\author{L. Perivolaropoulos}
\email{http://leandros.physics.uoi.gr} \affiliation{Department of
Physics, University of Ioannina, Greece}
\date{\today}

\begin{abstract}
We consider a universe with a compact extra dimension and a
cosmological constant emerging from a suitable ultraviolet cutoff on
the zero point energy of the vacuum. We derive the Casimir force
between parallel conducting plates as a function of the following
scales: plate separation, radius of the extra dimension and cutoff
energy scale. We find that there are critical values of these scales
where the Casimir force between the plates changes sign. For the
cutoff energy scale required to reproduce the observed value of the
cosmological constant, we find that the Casimir force changes sign
and becomes repulsive for plate separations less than a critical
separation $d_0=0.6mm$, assuming a zero radius of the extra
dimension (no extra dimension). This prediction contradicts Casimir
experiments which indicate an attractive force down to plate
separations of $100nm$. For a non-zero extra dimension radius, the
critical separation $d_0$ gets even larger than $0.6mm$ and remains
inconsistent with Casimir force experiments. We conclude that with
or without the presence of a compact extra dimension, vacuum energy
with any suitable cutoff can not play the role of the cosmological
constant.
\end{abstract}

\pacs{98.80.Es,98.65.Dx,98.62.Sb}
\maketitle

Cosmological observations during the last decade have indicated that
the universe expansion has been
accelerating\cite{Riess:1998cb,Davis:2007na} in contrast to
expectations based on the attractive gravitational force of matter
that would tend to decelerate the universe expansion. The observed
accelerating expansion of the universe indicates that either the
universe is dominated by an energy form with repulsive gravitational
properties (dark energy) or general relativity needs to be modified
on cosmological scales (modified gravity) (see eg
\cite{Copeland:2006wr} and references therein). Dark energy may
appear in the form of a cosmological constant (constant energy
density \cite{Padmanabhan:2002ji}), quintessence (variable energy
density due to an evolving scalar field \cite{Copeland:2006wr}), a
perfect fluid filling throughout space \cite{Gorini:2002kf} etc. The
simplest among the above possibilities is the cosmological constant
which is also favored by most cosmological data \cite{Serra:2007id}
compared to alternative more complicated models. It is therefore
important to identify the possible origins of such a cosmological
constant.

There have been several proposals concerning the origin of the
cosmological constant even though none of them is completely
satisfactory. Some of them include the zero point energy of the
vacuum \cite{Padmanabhan:2004qc}, anthropic considerations
\cite{Garriga:2005ee}, brane cosmology \cite{Binetruy:2000wn},
degenerate vacua \cite{Yokoyama:2001ez} etc. The simplest physical
mechanism that could lead to a cosmological constant is the
zero-point energy of the vacuum made finite by an ultraviolet
cutoff. A natural value of such a cutoff is the Planck scale leading
to an energy density of the vacuum $\rho_V=(2.44\times 10^{27})^4
eV^4$. However, such a value of the energy density leads to a
cosmological constant which is 123 orders of magnitude larger than
the observed one.

If the accelerating expansion of the universe is due to a
cosmological constant emerging due to zero point fluctuations of the
vacuum, the required energy density of the vacuum would be
$\rho_\Lambda \simeq 10^{-11} eV^4$ corresponding to a cutoff scale
(see eq. (\ref{vacen1}) below) of $\omega_c\simeq 10^{-3}eV$
($l_c\simeq 0.1 mm$). Even though there is no apparent physical
motivation for such a cutoff scale, it can not be a priori excluded
unless it is shown to be in conflict with specific experimental
data.

There are various types of laboratory experiments which are able to
probe directly quantities related to the zero point energy of the
vacuum. Such experiments include measurements of the Casimir force
\cite{expts} and measurements of the current noise in Josephson
junctions \cite{Jetzer:2006pt}. According to the Casimir
effect\cite{cas,Milton:2004ya,Milton:2000av}, the presence of
macroscopic bodies (like a pair of conducting plates) leads to a
modification of the zero point vacuum fluctuations due to the
introduction of non-trivial boundary conditions. This modification
manifests itself as a force between the macroscopic bodies that
distort the vacuum. When the role of the macroscopic bodies is
played by a pair of conducting plates, the discreteness imposed on
the fluctuation field modes, lowers the vacuum energy and leads to
an attractive force between the plates (the {\it Casimir force})
which has been measured by several experiments \cite{expts}. In the
presence of a finite cutoff length scale\cite{minlenpap} $l_c$,
electromagnetic field modes with wavelengths $\lambda < l_c$ are
suppressed and the vacuum energy lowering due to the presence of the
plates gets modified. This in turn leads to modifications of the
Casimir force between the plates compared to the case where no
cutoff has been imposed. In fact it may be shown
\cite{simid,Doran:2006ki,Mahajan:2006mw} that when the plate
separation $d$ becomes significantly less than the cutoff scale
$l_c$ then the force between the plates changes sign and becomes
repulsive! However, there is no experimental indication for change
of sign of the Casimir force down to separations $d\simeq 100nm$.
This imposes a constraint on the vacuum energy cutoff as $l_c <
100nm$ corresponding to a vacuum energy $\rho_V > 1 eV^4$. Such a
value of $\rho_V$ is clearly inconsistent with the vacuum energy
corresponding to the cosmological constant ($\rho_\Lambda \simeq
10^{-11} eV^4$). This inconsistency increases further if the
contribution of interactions other that electromagnetism is included
since in that case the predicted $\rho_V$ from the Casimir effect
would get even larger to include the contribution of other fields.
\begin{figure}[!t]
\rotatebox{0}{\hspace{0cm}\resizebox{0.45\textwidth}{!}{\includegraphics{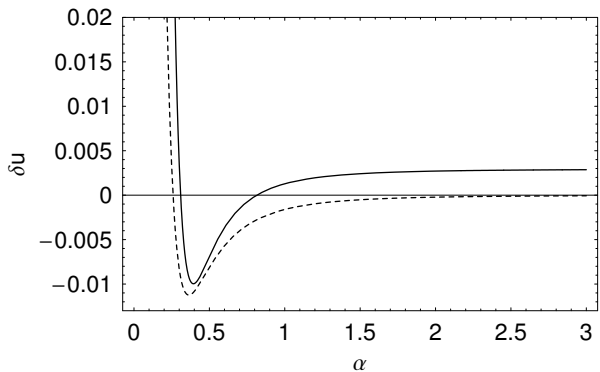}}}
{\hspace{0pt}\caption{The normalized vacuum energy ${\delta u}\equiv
\frac{\Delta u}{(\frac{3 \omega_{c}^3 \hbar}{4\pi c^2})}$ as a
function $\alpha$ (dimensionless form of $d$) for $\beta=0$ (dashed
line) and $\beta=1$ (continuous line). The curve corresponding to
$\beta=1$ has been shifted to lower values so that the locations of
the minima can be compared more easily.Notice that the minimum
shifts slightly to larger values as we increase the extra dimension
size.}} \label{fig1}
\end{figure}
It is therefore clear that either vacuum energy is not responsible
for the accelerating expansion of the universe or there is a missing
ingredient in the calculation of the Casimir force as a function of
the cutoff scale.

One such possible missing ingredient could be the presence of a
universal compact extra dimension \cite{casextdim} with
compactification scale $R$. Even though the current experimental
experimental bounds on $R$ are quite stringent ($R\lsim
(300GeV)^{-1}\simeq 10^{-9}nm$ \cite{Appelquist:2000nn}) it is still
instructive to consider this possibility in the context of the
Casimir effect with a cutoff in the vacuum energy. The propagation
of vacuum energy modes along this extra dimension modifies the field
spectrum and affects accordingly the Casimir force.

The main goal of this brief report is to address the following
question: `What is the effect of an extra compact dimension on the
Casimir force between two parallel plates and how does this effect
change as a function of the vacuum energy cutoff scale'.

\begin{figure}[!t]
\rotatebox{0}{\hspace{0cm}\resizebox{0.45\textwidth}{!}{\includegraphics{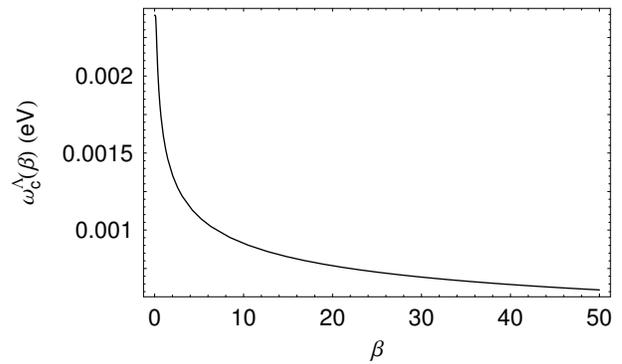}}}
{\hspace{0pt}\caption{The cutoff scale $\omega_c^\Lambda (\beta)$
required so that zero point vacuum fluctuations are identified with
the observed cosmological constant.  Notice that the cutoff scale
decreases as the extra dimension size increases.}} \label{fig2}
\end{figure}
We start by deriving the vacuum energy in the region between two
parallel plates. Consider a 3 dimensional Euclidean space with an
extra compact dimension of scale $R$. In this space consider a box
of square parallel conducting plates of large surface $L^2$ parallel
to the $xy$ plane placed at a small distance $d$ apart. The vacuum
energy of the quantized electromagnetic field in the region between
the plates \cite{cas,martin,casextdim} is \be {\cal E}=\frac{1}{2}
\sum_{{\vec k},\lambda}{\hbar \omega_{\vec k}}\;
g(\frac{\omega_{\vec k}}{\omega_c}) \label{vacen1} \ee where $g(x)$
is a UV cutoff regulator ($\lim_{x\rightarrow 0}g(x)=1$,
$\lim_{x\rightarrow \infty}g(x)=0$) and  $\lambda$ counts the
polarization modes.

Also \be \omega_{\vec k}=c\sqrt{k_x^2 + k_y^2 +k_d^2 + k_R^2}
\label{omk} \ee where \be k_x=\frac{\pi m_x}{L}\;\;k_y=\frac{\pi
m_y}{L}\;\;k_d=\frac{\pi n}{d}\;\;k_R=\frac{N}{R} \label{kmL} \ee
with $m_x.m_y,n,N=0,1,2...$. In the large $L$ limit eq.
(\ref{vacen1}) takes the form \be \frac{\cal E}{L^2}=\frac{3 \hbar
c}{4\pi} {\sum_{n,N=0}^\infty}^\prime \int_0^\infty dq \;q\;
\omega_q \; g(\frac{\omega_{\vec k}}{\omega_c}) \label{vacen2} \ee
where $\omega_q^2=c^2 \sqrt{q^2+(\frac{\pi
n}{d})^2+(\frac{N}{R})^2}$ and the factor 3 is due to the three
polarization degrees of freedom in the presence of the extra
dimension.
\begin{figure*}[!t]
\begin{minipage}{18pc}
\hspace{0cm}\rotatebox{0}{\resizebox{1\textwidth}{!}{\includegraphics{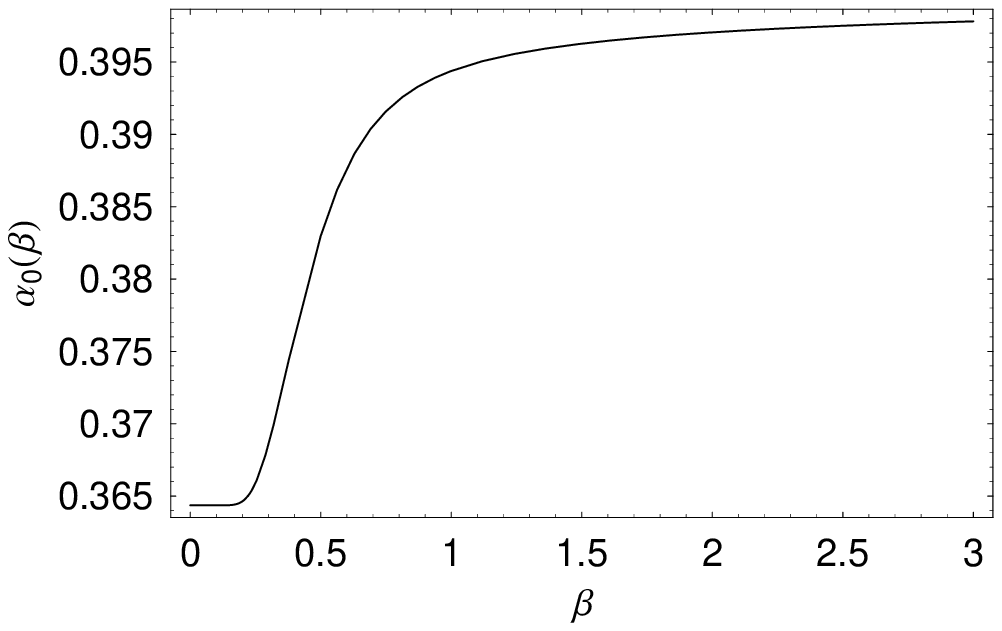}}}
\end{minipage}\hspace{4pc}
\begin{minipage}{18pc}
\hspace{0cm}\vspace{0.2cm}\rotatebox{0}{\resizebox{1\textwidth}{!}{\includegraphics{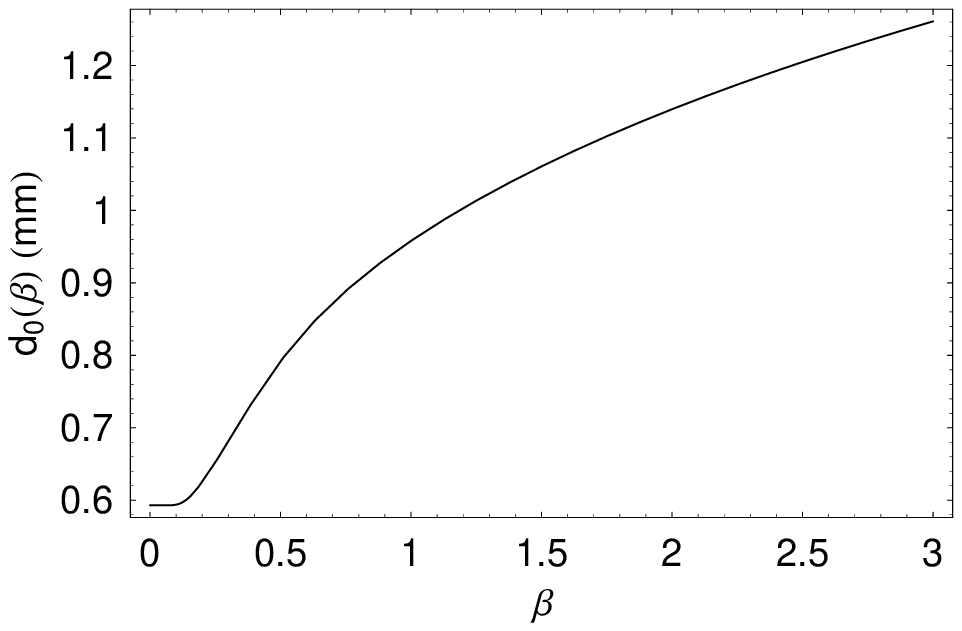}}}
\end{minipage}\hspace{4pc}
\vspace{0pt}{\caption{ The plate distance (dimensionless
$\alpha_0(\beta)$ (a), dimensionful $d_0(\beta)$ in $mm$ (b)) for
which the Casimir force changes sign, as a function of the
dimensionless scale of compatification $\beta$. In converting from
the dimensionless distance $\alpha_0(\beta)$ to the dimensionful
distance $d_0(\beta)$ in $mm$ we have assumed that the cutoff is
equal to $\omega_0^\Lambda (\beta)$ (eq. (\ref{omlc})) required to
reproduce the observed cosmological constant with vacuum energy.}}
\label{fig3}
\end{figure*}
The prime (') on the sum implies that when $N=n=0$ we should put a
factor $\frac{1}{3}$ (only one degree of freedom from polarization)
while if only one of the $N$, $n$ is 0 we should put an extra factor
$\frac{2}{3}$ (only two degrees of freedom from polarization).
Changing variable in eq. (\ref{vacen2}) we find the vacuum energy
per unit surface as: \be u^{vac}(d,R)=\frac{\cal E}{L^2}=\frac{3
\omega_c^3 \hbar}{4\pi c^2}
{\sum_{N,n=0}^{\infty}}^\prime\int_{0}^{\infty}\int_{\sqrt{(\frac{n}{\alpha})^2+(\frac{N}{\beta})^2}}^\infty
dv\; v^2 g(v) \label{uvac1} \ee where \be \alpha \equiv
\frac{\omega_c d}{c \pi} \;\;\;\; \beta \equiv \frac{\omega_c R}{c}
\label{abdef} \ee are the dimensionless forms of $d$ and $R$
respectively.
\begin{figure}[!t]
\rotatebox{0}{\hspace{0cm}\resizebox{0.45\textwidth}{!}{\includegraphics{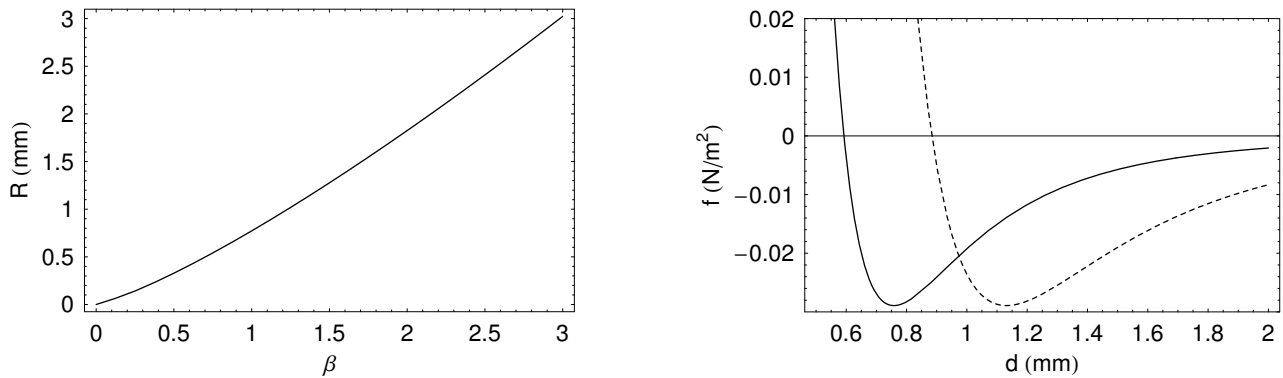}}}
{\hspace{0pt}\caption{The conversion of the dimensionless parameter
$\beta$ to the compactification scale $R$ for a cosmological
constant cutoff.}} \label{fig4}
\end{figure}
The corresponding result for large plate separation is \be
u_{>>}^{vac}(d,R)=\frac{3 \omega_{c}^3 \hbar}{4\pi
c^2}{\sum_{N=0}^{\infty}}^\prime \int_0^{\infty}dq
\int_{\sqrt{(\frac{q}{\alpha})^2+(\frac{N}{\beta})^2}}^\infty dv\;
v^2 g(v) \label{uvacll} \ee where the prime (') implies that a
factor $\frac{2}{3}$ should be included for $N=0$ to take proper
account of the number of degrees of freedom. Thus, the modification
of the vacuum energy due to the presence of the plates
is \ba && \Delta u(d,R)\equiv u^{vac}(d,R)-u_{>>}^{vac}(d,R)=\nn \\
&&=\frac{3 \omega_c^3 \hbar}{4\pi c^2}\sum_{N=0}^\infty
\left(\sum_{n=0}^\infty F(n,N) - \int_0^\infty dq F(q,N) \right)-\nn
\\ &&-h(n,N) \label{dudef} \ea where \ba && h(n,N)=\frac{3
\omega_{c}^3 \hbar}{4\pi c^2}[\frac{2}{3} F(0,0)+ \frac{1}{3}
\sum_{n=1}^\infty F(n,0) + \nn \\ && + \frac{1}{3} \sum_{N=1}^\infty
F(0,N) - \frac{1}{3} \int_0^\infty dq\; F(q,0)] \label{hdef}\ea and
\be F(n,N)\equiv
\int_{\sqrt{(\frac{n}{\alpha})^2+(\frac{N}{\beta})^2}}^\infty  dv \;
v^2\; g(v) \label{fnndef} \ee In the limit $\beta \rightarrow 0$
corresponding to $R\rightarrow 0$ (no extra dimension), only the
$N=0$ mode contributes and the above expressions reduce to the well
known \cite{martin,Mahajan:2006mw} forms of the regularized vacuum
energy.

\begin{figure}[!t]
\rotatebox{0}{\hspace{0cm}\resizebox{0.45\textwidth}{!}{\includegraphics{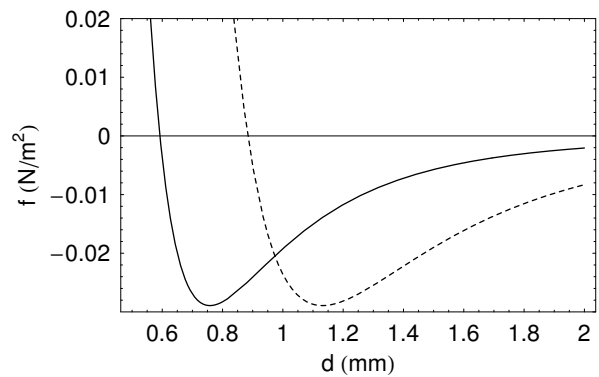}}}
{\hspace{0pt}\caption{The Casimir force per unit surface predicted
for a vacuum cutoff corresponding to the cosmological constant as a
function of the  plate distance $d$ in $mm$ for $\beta=0$ (no extra
dimension, continuous line) and $\beta=1$ (dashed line).}}
\label{fig5}
\end{figure}

We now specify an exponential form for the UV cutoff function
$g(v)$. Other smooth forms of $g(v)$ also lead to similar results.
For $g(v)=e^{-v}$ we have $F(n,N)=e^{-s}(2+s(s+2))$ with
$s=\sqrt{(\frac{n}{\alpha})^2+(\frac{N}{\beta})^2}$. In Fig. 1 we
show the normalized vacuum energy ${\delta u}\equiv \frac{\Delta
u}{(\frac{3 \omega_{c}^3 \hbar}{4\pi c^2})}$ as a function of
$\alpha$  for $\beta=0$ and $\beta=1$. Notice that for a small value
of $R$ ($R\omega_c \rightarrow 0$) we recover the result of ref.
\cite{Mahajan:2006mw} where the repulsive nature of the Casimir
force was demonstrated for plate separations $d$ much smaller than
the cutoff scale $c \omega_c^{-1}$ without the presence of extra
dimensions. For $\beta=0$ (no extra dimension), the Casimir force
becomes repulsive for $\alpha<\alpha_0 \simeq 0.36$. For a cutoff
leading to the observed value of cosmological constant ($\omega_c
\simeq 10^{-3} eV$) we find the critical separation $d_0 \simeq
0.6mm$ such that for separations $d<d_0$ the Casimir force becomes
repulsive. Since the Casimir force has been experimentally shown to
be attractive down to plate separations $d\simeq 100nm$ \cite{expts}
it becomes clear that the observed cosmological constant can not be
due to vacuum energy with appropriate cutoff and no extra
dimensions.

The introduction of a compact extra dimension with finite size has
two effects:
\begin{itemize} \item The cutoff scale $\omega_c(\beta)$ required to
match the observed value of the cosmological constant slowly
decreases \item The critical dimensionless separation $\alpha_0
(\beta)$ for which the Casimir force changes sign increases (see
Fig. 1). \end{itemize}

In order to demonstrate the first effect we may evaluate the
predicted vacuum energy density as a function of the dimensionless
size $\beta$ of the extra dimension. We have \be \rho_V (\omega_c,
\beta)=\frac{u_{>>}^{vac}(d)}{d}=\frac{3 \hbar \omega_c^4}{4\pi^2
c^3} Q(\beta) \label{rhovom} \ee where \be Q(\beta)=
{\sum_{N=0}^\infty}^\prime\int_0^\infty dq' \int_{\sqrt{q'^2 +
(\frac{N}{\beta})^2}}^\infty v^2 e^{-v} \ee Demanding \be \rho_V
(\omega_c,\beta)=\rho_\Lambda = 10^{-11} eV^4 \label{rvrl} \ee we
find \be \omega_c^\Lambda (\beta)=(\frac{4 \pi^2}{3} Q(\beta)^{-1}
10^{-11})^{1/4} eV \label{omlc} \ee A plot of $\omega_c^\Lambda
(\beta)$ is shown in Fig. 2. In Fig. 3 we show a plot of the
dimensionful plate distance $d_0 (\beta)$ corresponding to a sign
change of the Casimir force. This is found by converting the
dimensionless minima $\alpha_0$ of Fig. 1 to the corresponding
dimensionful plate distances $d_0$ assuming a cutoff equal to
$\omega_c^\Lambda(\beta)$ in eq. (\ref{abdef}) for $\alpha$. As
shown in Fig. 3 $d_0(\beta)$  slowly increases with $\beta$ as
expected by inspection of Fig. 1 which shows the minimum $\alpha_0$
of $\Delta u (\alpha, \beta)$ shifts to larger values as we increase
the dimensionless size $\beta$ of the extra dimension.

It is easy to convert the dimensionless parameter $\beta$ to the
compactification scale $R$ for a particular value of the cutoff
$\omega_c$. For example for the cosmological constant cutoff
$\omega_c=\omega_c^\Lambda (\beta)$ we obtain using eqs.
(\ref{abdef}), (\ref{omlc}) \be R=\frac{3\beta}{(8.2\times
Q(\beta))^{-1/4}} \label{rbconv} \ee which is shown in Fig. 4.

Finally, it is straightforward to use eqs (\ref{dudef}),
(\ref{abdef}) to evaluate the Casimir force for
$\omega_c=\omega_c^\Lambda (\beta)$. The resulting force is shown in
Fig. 5 for $\beta=0$ ($R=0$) and $\beta=1$ ($R\simeq 0.7mm$).
Clearly, the plate separation where the force changes sign depends
weakly on the size of the extra dimension and is always larger than
$0.6mm$. We conclude that even in the presence of compact extra
dimensions a cosmological constant induced by zero point vacuum
fluctuations with appropriate cutoff is in conflict with
experimental measurements of the Casimir force which indicate an
attractive force down to separations of $d\simeq 100nm$.

Finally, we point out that our approach connecting the Casimir
effect in the presence of a cutoff with the cosmological constant is
distinct from  previous global approaches \cite{Elizalde:2000jv}
which view the universe as a system of large (cosmological size) and
small compact dimensions (finite system) where the vacuum energy
gets modified due to the finiteness of the system. In these models a
cosmological constant is generated by the Casimir energy associated
to some field propagating in the extra dimensions. That mechanism
which does not implement a cutoff and does not address the issue of
ultraviolet contributions (set to zero by an unknown mechanism),
remains a valid candidate model for the generation of a cosmological
constant but it can not be tested with Casimir force laboratory
experiments even though tight constraints are imposed by the
predicted deviations from Newton's law.

This work was supported by the European Research and Training
Network MRTPN-CT-2006 035863-1 (UniverseNet).

\end{document}